\documentclass[prx,twocolumn,superscriptaddress]{revtex4}

\usepackage{graphicx,amsmath}
\usepackage{feynmp}
\usepackage{subfigure}
\usepackage{epstopdf}

\usepackage{color}

\begin{document}

\title{Majorana Zero Modes and Topological Quantum Computation}
\author{Sankar Das Sarma}
\affiliation{Department of Physics, University of Maryland,
College Park, MD 20742}
\affiliation{Microsoft Station Q, University of
California, Santa Barbara, CA 93108}
\author{Michael Freedman}
\affiliation{Microsoft Station Q, University of
California, Santa Barbara, CA 93108}
\author{Chetan Nayak}
\affiliation{Microsoft Station Q, University of
California, Santa Barbara, CA 93108}
\affiliation{Department of Physics, University of California, Santa Barbara, California 93106, USA}

\begin{abstract}
We provide a current perspective on the rapidly developing field of Majorana zero modes in solid state systems.
We emphasize the theoretical prediction, experimental realization, and potential use of Majorana zero modes
in future information processing devices through braiding-based topological quantum computation.
Well-separated Majorana zero modes should manifest non-Abelian braiding statistics suitable for unitary gate operations for topological quantum computation.
Recent experimental work, following earlier theoretical predictions, has shown specific signatures consistent with the existence of Majorana modes localized
at the ends of semiconductor nanowires in the presence of superconducting proximity effect.  We discuss the experimental findings and their theoretical analyses,
and provide a perspective on the extent to which the observations indicate the existence of anyonic Majorana zero modes in solid state systems. 
We also discuss fractional quantum Hall systems (the 5/2 state), which have been extensively studied in the context of non-Abelian anyons and topological quantum computation..  We describe proposed schemes
for carrying out braiding with Majorana zero modes as well as the necessary steps for implementing topological quantum computation.
\end{abstract}

\maketitle

\section{Introduction}

Topological quantum computation \cite{Kitaev97,Freedman98}, is an approach to fault-tolerant quantum computation in which
the unitary quantum gates result from the braiding of certain topological quantum objects, called `anyons'.
Anyons braid nontrivially: two counter-clockwise exchanges do not leave the state of the system invariant,
unlike in the cases of bosons or fermions. Anyons can arise in
two ways: as localized excitations of an interacting quantum Hamiltonian \cite{Nayak08} or as defects in an ordered system
\cite{Etingof10,Barkeshli14}. Fractionally-charged excitations of the Laughlin fractional quantum Hall liquid
are an example of the former. Abrikosov vortices in a topological superconductor are an example of the latter.
Not all anyons are directly useful in topological quantum computation; only non-Abelian anyons are useful,
which does not include the anyonic excitations (sometimes referred to as Abelian anyons, to distinguish them from the more exotic non-Abelian anyons which are useful for topological quantum computation) that are believed to occur in most odd-denominator
fractional quantum Hall states. A collection of non-Abelian anyons at fixed positions and with fixed local quantum numbers
has a non-trivial topological degeneracy (which is, therefore, robust -- i.e. immune to weak local perturbations).  This topological degeneracy allows quantum computation since braiding enables unitary operations between the distinct degenerate states of the system. The unitary transformations resulting from braiding depend only
on the topological class of the braid, thereby endowing them with fault-tolerance. This topological immunity is protected
by an energy gap in the system and a length scale discussed below.  As long as the braiding operations are slow compared with the inverse of the energy gap and external perturbations are not strong enough to close the gap, the system remains robust to disturbances and noise. These braiding operations constitute the elementary gate operations for the evolution of the topological quantum computation.

Perhaps the simplest realization of a non-Abelian anyon is a quasiparticle
or defect supporting a Majorana zero mode (MZM). (The zero-mode here refers to the zero-energy
midgap excitations that these localized quasiparticles typically correspond to in a low-dimensional topological superconductor.)
This is a real fermionic operator that commutes with the Hamiltonian. The existence of such operators guarantees
topological degeneracy and, as we explain in Section \ref{sec:what-is-MZM}, braiding necessarily causes
non-commuting unitary transformations to act on this degenerate subspace. The term ``Majorana''
refers to the fact that these fermion operators are real, as in Majorana's real version of the Dirac equation.
However, there is little connection with Majorana's original work or its application to neutrinos. 
Rather, the key concept here is the non-Abelian anyon, and MZMs are a particular mechanism by which
a particular type of non-Abelian anyons, usually called ``Ising anyons'' can arise.
By contrast, Majorana fermions, as originally conceived, obey ordinary Fermi-Dirac statistics, and are simply a particular type of fermion.
Although the terminology `Majorana fermions' is somewhat misleading for MZMs, it is used extensively in the literature.

If Majorana zero modes can be manipulated and their states measured in well-controlled experiments, this could pave the way
towards the realization of a topological quantum computer.  The subject got a tremendous boost in 2012 when an experimental group in Delft published evidence for the existence of Majorana zero modes in InSb nanowires \cite{Mourik12}, following
earlier theoretical predictions \cite{Lutchyn10,Oreg10,Sau10b}.
The specific experimental finding, which has been reproduced later in other laboratories, is a zero-bias tunneling conductance peak in a semiconductor (InSb or InAs) nanowire in contact with an ordinary metallic superconductor (Al or Nb), which shows up only when a finite external magnetic field is applied to the wire.
Several other experimental groups also saw evidence (i.e. zero bias tunneling conductance peak in an applied magnetic field) for the existence of Majorana zero-modes in both InSb and InAs nanowires
\cite{Rokhinson12,Deng12,Churchill13,Das12,Finck12},
thus verifying the Delft finding.  However, though these experiments are compelling,
they do not show exponential localization with system length required
by Eq. ({eqn:MZM-real-def}) or anyonic braiding behavior.
As explained later in this article, the exponential localization of the isolated Majorana modes at wire ends and the associated non-Abelian braiding properties are the key features which enable topological quantum computation to be possible in these systems.

In the current article, we provide a perspective on where this interesting and important subject is today (at the end of 2014).  This is by no means a review article for the field of Majorana zero modes or the topic of topological quantum computation since such reviews will be too lengthy and too technical for a general readership.  There are, in fact, several specialized review articles already discussing various aspects of the subject matter which we mention here for the interested reader.  The subject of topological quantum computation has been reviewed by us in great length earlier \cite{Nayak08}, and we have also written a shorter version of anyonic braiding-based topological quantum computation elsewhere \cite{DasSarma06b}.  There are also several excellent popular articles on the braiding of non-Abelian anyons and topological quantum computation
\cite{Collins06,Venema08}. The theory of Majorana zero-modes and their potential application to topological quantum computation
has recently been reviewed in great technical depth in several articles \cite{Alicea12a,Leijnse12,Beenakker2013a,Stanescu13}.

There are essentially two distinct physical systems that have been primarily studied in the search for Majorana zero modes for topological quantum computation (TQC).  The first is the so-called 5/2-fractional quantum Hall system (“5/2-FQHS”) where the application of a strong perpendicular magnetic field to a very high-mobility two-dimensional (2DEG) electron gas (confined in epitaxially-grown GaAs-AlGaAs quantum wells) leads to the even-denominator fractional quantization of the Hall resistance. The generic fractional quantum Hall effect leads to the quantization with odd-denominator fractions (e.g. the original $1/3$ quantization observed in the famous experiment by Tsui, Stormer, and Gossard in 1982 \cite{Tsui82}).  Interestingly, of the almost 100 FQHS states that have so far been observed in the laboratory, the 5/2-FQHS is the only even-denominator state ever found in a single 2D layer.
It has been hypothesized that this even-denominator state supports Ising anyons. A topological qubit was proposed by us for this platform
\cite{DasSarma05} in 2005, building upon previous theoretical work on the
$5/2$ state \cite{Moore91,Greiter92,Nayak96c,Fradkin98,Morf98}. Tantalizing experimental signatures for the possible existence of the desired non-Abelian anyonic properties were reported in subsequent experiments \cite{Willett09,Willett10,Willett13a,Willett13b}.
However, these results have not been reproduced in other laboratories. Potential barriers to progress are 
the required extreme high sample quality (mobility $>10^7$ cm$^2$/V.s), very low $< 25$mK temperature
and high magnetic field $>2T$. 
The second system is the semiconductor nanowire structure proposed in Refs. \cite{Lutchyn10,Oreg10},
building upon earlier theoretical work on topological superconductors
\cite{Read00,Kitaev01,Fu08,Sau10a}.
Semiconductor nanowires are the focus of this paper, but the $5/2$ fractional quantum Hall state is a useful
point of comparison since a great deal of experimental and theoretical work has been done on the 5/2 FQHS over the last 27 years.

\section{What is a Majorana Zero Mode?}
\label{sec:what-is-MZM}

A Majorana zero mode (MZM)  is a fermionic operator $\gamma$ that squares to $1$ (and, therefore, is necesarily self-adjoint)
and commutes with the Hamiltonian $H$ of a system:
\begin{equation}
\label{eqn:MZM-ideal-def}
\gamma \mbox{ fermionic} \, , \,\, \gamma^2 = 1 \, , \,\, [H,\gamma] = 0
\end{equation}
Any operator that satisfies the first two conditions is called a Majorana fermion operator.
If it satisfies the third condition, as well, then it is a Majorana zero mode operator or, simply,
a Majorana zero mode\cite{Note1}.
The existence of such operators implies the existence of a degenerate space of ground states, in which quantum information can be stored. If there are $2n$ Majorana zero modes, $\gamma_{1}, \ldots \gamma_{2n}$ (they must come in pairs since each MZM
is, in a sense, half a fermion) satisfying
\begin{equation}
\label{eqn:CAR}
\{ \gamma_{i} , \gamma_{j} \} = 2\delta_{ij}
\end{equation}
then the Hamiltonian can be simultaneously diagonalized with the operators $i\gamma_{1} \gamma_{2}$, $i\gamma_{3} \gamma_{4}$, $\ldots$, $i\gamma_{2n-1} \gamma_{2n}$. The ground states can be labelled by the eigenvalues $\pm 1$ of these $n$ operators, thereby leading to a $2^n$-fold degeneracy. There is a two-state
system associated with each {\it pair} of MZMs. This is to be contrasted with a collection of spin-$1/2$ particles,
for which there is a two-state system associated with each spin. In the case of MZMs, we are free to pair them
however we like; different pairings correspond to different choices of basis in the $2^n$-dimensional ground
state Hilbert space.

Unfortunately, the preceding mathematics is too idealized for a real physical system.
If we are fortunate, there can, instead, be self-adjoint Majorana fermion
operators $\gamma_{1}, \ldots \gamma_{2n}$ satisfying the anti-commutation relations (\ref{eqn:CAR}) and
\begin{equation}
\label{eqn:MZM-real-def}
[H, \gamma_{i}] \sim e^{-x/\xi}
\end{equation}
where $x$ is a length scale mentioned in the introduction (which can be construed to be
the separation between two MZMs in the pair) and discussed
momentarily, and $\xi$ is a correlation length associated with the Hamiltonian $H$.
In the superconducting systems that will be discussed in the sections to follow, $\xi$ will be the superconducting coherence length.
All states above the $2^{n-1}$-dimensional low-energy subspace have a minimum energy $\Delta$.
In order for the definition (\ref{eqn:MZM-real-def}) to approach the ideal condition
(\ref{eqn:MZM-ideal-def}), it must be possible to make $x$ sufficiently large that the right-hand-side of
Eq. (\ref{eqn:MZM-real-def}) approaches zero rapidly.
This can occur if the operators $\gamma_i$ are localized at points $x_i$
(which we have not, so far, assumed). Then
$\gamma_i$ commutes or anti-commutes,
up to corrections $\sim e^{-y/\xi}$, with, respectively, all
local bosonic or fermionic operators that can be written
in terms of electron creation and annihilation operators
whose support is a minimum distance $y$ from some point $x_i$.
The effective Hamiltonian for energies much lower than $\Delta$
is a sum of local terms, which means that products of operators such as $i \gamma_{i} \gamma_{j}$
must have exponentially-small coefficients $\sim e^{-|{x_i}-{x_j}|/\xi}$ \cite{Note2}.
Consequently, the condition
(\ref{eqn:MZM-real-def}) then holds \cite{Note4}.
The number of Majorana zero mode operators satisfying (\ref{eqn:MZM-real-def}) must be even.
Consequently, if we add a term to the Hamiltonian that couples a single zero mode operator to the non-zero mode operators,
a zero mode operator will remain since zero modes can only be lifted in pairs.
Thus, the exponential `protection' of the MZMs allowing their quantum degeneracy  is enabled by the energy gap, which should be as large as possible for effective TQC operations. Thus, in a loose sense, two Majoranas together give a Dirac fermion, and these two MZMs must be far away from each other for the exponential topological protection to apply.

It is useful to combine the
two MZMs into a single Dirac fermion $c = \gamma_1 + i\gamma_2$. The two states of this pair
of zero modes corresponds to the fermion parities $c^\dagger c = 0, 1$.
Thus, if the total fermion
parity of a system is fixed, then the degeneracy of $2n$ MZMs is $2^{n-1}$-fold.
This quantum degeneracy, arising from the topological nature of the MZMs, enables TQC to be feasible by braiding the MZMs around each other.

Such localized MZMs are known to occur in two related but distinct physical situations.
The first is at a defect in an ordered state, such as a vortex in a superconductor or a domain wall
in a 1D system.
The defect does not have finite energy in the thermodynamic limit and, therefore,
it is not possible to excite a pair of such defects at finite energy cost and pull them apart. However,
by tuning experimental parameters (which involves energies proportional to the system size),
such defects can be created in pairs, thereby creating pairs of MZMs. The best example of this
is a topological superconductor.
Alternatively, there may
be finite-energy quasiparticle excitations of a topological phase \cite{Nayak08}
that support zero modes. This scenario is believed to
be realized in the $\nu=5/2$ fractional quantum Hall states, where charge $e/4$ excitations
are hypothesized to support MZMs. Although the cases of defects in topological supercondcutors
and quasiparticles in "true" topological phases are closely-related,
there are some important differences, touched on later.

When two defects or quasiparticles supporting MZMs are exchanged while maintaining
a distance greater than $\xi$, their MZMs must also be exchanged.
Since the $\gamma_i$ operators are real, the exchange process can, at most, change their signs.
Moreover, fermion parity must be conserved, which dictates that $\gamma_1$ and $\gamma_2$ must
pick up opposite signs. Hence, the transformation law is:
\begin{equation}
\gamma_1 \rightarrow \pm \gamma_2 \, \, , \,\,\, \gamma_2 \rightarrow \mp \gamma_1
\end{equation}
The overall sign is a gauge choice. This transformation is generated by the unitary operator:
\begin{equation}
\label{eqn:braiding-unitary}
U = e^{i\theta} \, e^{\frac{\pi}{4}\gamma_{1}\gamma_{2}}
\end{equation}
This is the braiding transformation of Ising anyons. Strictly speaking, Ising anyons have $\theta=\pi/8$.
Other values of $\theta$ can occur if there are additional Abelian anyons attached to the Ising anyons,
as is believed to occur in the $\nu=5/2$ fractional quantum Hall state. In the case of defects, rather
than quasiparticles, the phase $\theta$ will not, in general, be universal, and will depend on the
particular path through which the defects were exchanged. We emphasize that this braiding
transformation law follows from (a) the reality condition of the Majorana fermion operators $\gamma_{1,2}$,
(b) the locality of the MZMs, and (c) conservation of fermion parity. Therefore, an experimental observation
consistent with such a braiding transformation is evidence that (a)-(c) hold. This, in turn is
evidence that the defects or quasiparticles support Majorana zero modes satisfying
the definition (\ref{eqn:MZM-real-def}). Such a direct experimental observation of braiding has not yet
happened in the laboratory.

In the case of quasiparticles in topological phases, braiding properties, as revealed through
various concrete proposed interference experiments
such as those proposed in Refs. \onlinecite{Fradkin98,DasSarma05,Bonderson06a,Stern06}, is,
perhaps, the gold standard for detecting MZMs.
However, in the case of defects in ordered states and, in particular, in the
special case of MZMs in superconductors, a zero-bias peak in transport with a normal lead \cite{Sengupta01}
and a $4\pi$ periodic Josephson effect \cite{Kitaev01} are also signatures,
as discussed in Section \ref{sec:other-signatures}. Before discussing these in
more detail in Section \ref{sec:other-signatures}, it may be helpful to discuss the differences
between topological superconductors and true topological phases.

\section{Majorana Zero Modes in Topological Phases and in Topological Superconductors}
\label{sec:MZMs-in-SC}

As noted in the Introduction, Ising anyons can be understood as quasiparticles or defects that support Majorana zero modes.
In the Moore-Read Pfaffian state \cite{Moore91,Greiter92} and the anti-Pfaffian state \cite{LeeSS07,Levin07},
proposed as candidate non-Abelian states for the 5/2 FQHS,
charge-$e/4$ quasiparticles are Ising
anyons \cite{Nayak96c,Read96,Tserkovnyak03,Seidel08,Read08,Baraban09,Prodan09,Bonderson11a}.
There is theoretical~\cite{Morf98,Rezayi00,Feiguin08,Peterson08,Feiguin09,Bishara09a,Rezayi09,Zaletel15,Pakrouski15}
and experimental~\cite{Radu08,Dolev08,Willett09,Willett10,Bid10,Venkatachalam11,Tiemann11,Stern12,Willett13a,Willett13b} evidence
that the $\nu=5/2$ fractional quantum Hall state is in one of these two universality classes.
However, there are also some experiments \cite{Stern10,Rhone11,Lin12,Baer14} that do not agree with this hypothesis.
The non-Abelian statistics of quasiparticles at $\nu=5/2$ has been reviewed in Ref. \onlinecite{Nayak08} and would require
a digression into the physics of the fractional quantum Hall effect. Hence, we do not elaborate on it here, other than to
note that Ising-type fractional quantum Hall states are very nearly topological phases, apart from some
deviations that are salient on higher-genus surfaces \cite{Bonderson13}. However, the electrical charge that
is attached to Ising anyons enables their detection through charge transport experiments \cite{Fradkin98,DasSarma05,Bonderson06a,Stern06}.
Ising anyons also occur in some lattice models of gapped, topologically-ordered spin liquids \cite{Levin05a,Kitaev06a}.
These are true topological phases in which the MZM operators are associated with finite-energy excitations of the system and do not have a local relation to the underlying spin operators, much less the electron operators, whose charge degree of freedom is gapped. This limits the types of effects (in comparison to the superconducting case)
that could break the topological degeneracy
implied by Eqs. \ref{eqn:MZM-ideal-def} and \ref{eqn:CAR}.

MZMs also occur at defects in certain types of superconductors that form a subset of the class
generally called ``topological superconductors'' \cite{Volovik99,Read00,Kitaev01}. We discuss these in general terms
in this section and then in the context of specific physical realizations in Section \ref{sec:synthetic}.

Topological phases have some topological features and some “ordinary” non-topological features.
However, the interplay between these two types of physics is even more central in topological
superconductors. This is both ``bad'' and ``good.'' It is “bad” if the nontopological features represent an
opportunity for error or lead to energy splittings that decohere desirable superpositions.
It is “good” when they allow a convenient coupling to conventional physics,
something we had better have available if we ever wish to measure the topological system.
In topological phases, there is a trivial tensor product situation in which the topological and the ”ordinary” degrees of freedom
do not talk to each other.  In this case, we do not have to worry that the latter induce errors in the former, but they also will not be useful in initializing or measuring the topological degrees of
freedom. (As always, in discussing topological physics, we regard effects that diminish exponentially with length, frequency, or temperature as unimportant. This is somewhat analogous to computer scientists classifying algorithms as polynomial time or slower. Clearly the power and even the constants can make a difference, but such a structural dichotomy is a useful starting point.) So, for example, if there are phonons in a system, their interaction with topological degrees of freedom causes a splitting of
the topological degeneracy that vanishes as $e^{-L/\xi}$ at zero temperature \cite{Bonderson13}, so
we would consider the system as essentially a tensor product, with the phonons in a separate factor.
However, a topological superconductor is not a true topological phase but, rather, following the
terminology of Ref. \onlinecite{Bonderson13} a “fermion parity protected quasi topological phase”.
The qualifier ``quasi'' permits the existence of benign gapless modes as discussed above.
With slightly more precision: an excitation is topological if its local density matrices cannot be produced to high fidelity by a local operator acting from one of the system's ground states. ``Quasi'' permits low-energy  excitations (below the gap) provided they are not ``topological''. These subgap excitations
surely do exist in real topological superconductors: there will be phonons and there will be gapless excitations
of the superconducting order parameter  - both are Goldstone modes of broken symmetries
(translation in the first case and U(1)-charge conservation in the second).
(The reader may wonder why the now-so-famous Higgs mechanism fails to gap the Goldstone mode of
broken U(1). The answer is the mismatch of dimensions, the gauge field roams 3-dimensional space
while the superconductor lives in either two or one dimension. In the former case, the interaction with the gauge
field causes superconducting phase fluctuations to have dispersion $\omega \sim \sqrt{q}$ while in the latter
case $\omega\sim q$. In a bulk 3D super conductor the gauge boson is indeed gapped out.)
The more serious caveat is “fermion parity protected”.  This is simultaneously a blessing and a curse for any project to compute with Majorana zero modes in superconductors.
The blessing is that the basis states of the topological qubit have this precise interpretation: fermion parity.  If we are willing to move into an unprotected regime to measure them,
MZMs can be brought together and their charge parity detected locally. Using more sophistication,
one could keep the MZMs at topological separation and exploit the Aharonov-Casher effect
to measure the charge parity encircled by a vortex. So this coupling will allow measurement by physics very well in hand. (It is less clear how to do this with, for instance, the computationally more powerful Fibonacci anyons \cite{Nayak08}.)
Measurement is crucial for processing quantum information with MZMs since the braid group representation
for Ising anyons is a rather modest finite group: beyond input and output, distillation of quantum states is needed \cite{Bravyi05},
and this is measurement intensive. 
The curse is quasi-particle-poisoning.  A nearby electron can enter the system
and be absorbed by a Majorana zero mode, thereby flipping the fermion parity -- i.e. flipping a qubit.
The electrons' charge is absorbed by the superconducting condensate.
This propensity of a topological superconductor to be poisoned (or equivalently, the fermion parity to flip in an uncontrolled manner) represents a salient distinction from the Moore-Read state proposed for the $\nu=5/2$ fractional quantum Hall state. In the Moore-Read state,
the vortices carry electric charge ($\pm e/4$) and fermions carry charge $0$ or $\pm 1/2$.
Consequently, there is an energy gap to bringing an electron from the outside into a $\nu=5/2$ FQHE fluid.
Its fermion parity can be absorbed by a Majorana zero mode (as in the case of a topologial superconductor),
but there is no condensate to absorb its charge; instead, four disjoint charge-$e/4$ quasiparticles must be created,
with their attendant energy cost.
It would be harder to poison a $\nu=5/2$ fluid but also harder to discern its state and the signatures
discussed in the next section are not available for non-Abelian FQHS states.
Thus, one must choose between potentially better protection (5/2 FQHE) or easier measurement (topological superconductor).

\section{Signatures of MZM{s} in Topological Superconductors}
\label{sec:other-signatures}

Due to the superconducting order parameter, it is possible for an electron to tunnel directly into a MZM
in a superconductor. Suppose there is a MZM $\gamma$ at the origin ${\bf x}=0$ in
a superconductor. Then, if we bring a metallic wire near the origin, electrons can tunnel from the
lead to the superconductor via a coupling of the form
\begin{equation}
\label{eqn:zm-lead-coupling}
H_{\rm tun} = \lambda \,c^\dagger(0)\,\, \gamma \, e^{-i\theta(0)/2} +
\lambda^* \gamma \,\,c(0)\,\, e^{i\theta(0)/2}
\end{equation}
where $c(0)$ is the electron annihilation operator in the lead. For simplicity, we have suppressed
the spin index, which is a straightforward notational choice if the supercondutor and the lead are both fully spin-polarized.
In the more generic case, the spin index must be handled with slightly more care. Here,
$\theta$ is the phase of the superconducting order parameter. Ordinarily, we would expect that it would
be impossible for an electron, which carries electrical charge, to tunnel into a Majorana zero mode,
which is neutral since $\gamma=\gamma^\dagger$. However, the superconducting condensate
(which is a condensate of Cooper pairs that breaks the U(1) charge conservation symmetry)
can accomodate electrical charge, thereby allowing this process, which is a form of Andreev reflection.
In the case of the Moore-Read
Pfaffian quantum Hall state, however, this is not possible.
In order for an electron to tunnel into an MZM, four charge-$e/4$ quasiparticles
must also be created in order to conserve electrical charge. This can only happen when the bias voltage exceeds
four times the charge gap.

In the case of a topological superconductor, the coupling (\ref{eqn:zm-lead-coupling}), which seems like
a drawback as compared to a topological phase, can actually be an advantage
since it opens up the possibility
of a simple way of detecting Majorana zero modes that does not involve braiding them. 
For at $T, V \ll \Delta$, the electrical conductivity from a 1D wire through a contact
described by Eq. (\ref{eqn:zm-lead-coupling}) takes the form \cite{Sengupta01,Law09,Fidkowski12,Lutchyn13}:
\begin{equation}
G(V,T) = \frac{2e^2}{h} \, h(T/V,T/{\Lambda^*})
\end{equation}
where $h(0,0)=1$ and $\Lambda^*$ is a crossover scale determined by the tunneling strength,
$\Lambda^* \sim \lambda^y$, where the exponent $y$ depends on the interaction strength in
the 1D normal wire so that $y=1/2$ for a wire with vanishing interactions.
At low voltage and low temperature, the conductivity is $2{e^2}/h$, indicative of perfect Andreev reflection: each electron
that impinges on the contact is reflected as a hole and charge $2e$ is absorbed by the topological superconductor. There is vanishing
amplitude for an electron to be scattered back normally.
Such a conductivity can occur for other reasons (see, e.g. \cite{LeeEJH12,Bagrets12}), but they
are non-generic and require some
special circumstances and can, in principle, be ruled out by further experiments.
Thus the observation of perfect Andreev reflection, with the associated quantized conductance at zero bias, robust to parameter changes,
is an indication of the presence of a Majorana zero mode. In Section \ref{sec:experiments}, we discuss
the extent to which this quantized tunneling conductance associated with the zero-energy midgap Majorana modes has actually been observed in experiments.

A second probe of Majorana zero modes that is special to topological superconductors is the
the so-called fractional Josephson effect. When two normal superconductors are in electrical contact,
separated by a thin insulator or a weak link, the dominant
coupling between them at low temperatures is
\begin{equation}
H = - J \cos\theta
\end{equation}
where $\theta$ is the difference in the phases of the order parameters of the two superconductors. It
is periodic in $\theta$ with period $2\pi$. The Josephson current is the derivative of this coupling
with respect to $\theta$; it, too, is periodic in $\theta$ with period $2\pi$.
The Josephson coupling is proportional to the square of the amplitude for
an electron to tunnel from one superconductor to the other, $J\propto t^2$. However,
when two topologial superconductors are in contact and there are MZMs on both sides
of the Josephson junction, the leading coupling is:
\begin{equation}
H = - it {\gamma_L} {\gamma_R} \cos(\theta/2)
\end{equation}
So long as $i{\gamma_L} {\gamma_R}=\pm 1$ remains fixed during the measurement, the Josephson
current now has period $4\pi$, rather than $2\pi$ as in nontopological superconductors.
An observation
of the $4\pi$ `fractional' Josephson effect in AC measurements would
be compelling evidence in favor of the existence of MZMs in a superconducting system.
However, if $i{\gamma_L} {\gamma_R}=\pm 1$ can vary in order to find the
minimum energy at each value of $\theta$, then it will flip when $\cos(\theta/2)$ changes sign. Consquently, the
current will have period $2\pi$. The value of $i{\gamma_L} {\gamma_R}=\pm 1$ can change if a fermion is absorbed
by one of the zero modes ${\gamma_L}$ or ${\gamma_R}$. Such a fermion may come from a localized low-energy state
or an out-of-equilibrium fermion excited above the supercondcuting gap. In order to use the Josephson effect to detect
MZMs, an AC measurement must be done at frequencies higher than the inverse of the time scale for such processes.

This can be done through the observation of Shapiro steps \cite{Rokhinson12}.
When an ordinary Josephson junction is subjected to
electromagnetic waves at frequency $\omega$, a DC voltage develops and passes through
a series of steps $V_{\rm DC} = n \frac{h}{2e} \omega$ as the current is increased. However, when there
are Majorana zero modes at the junction, then the $4\pi$ periodicity discussed above translates to Shapiro steps
$V_{\rm DC} = n \frac{h}{e} \omega$. In essence, charge transport across a junction with MZMs is due to charge $e$
rather than charge $2e$ objects, so the flux periodicity and voltage steps are doubled.
In terms of conventional Shapiro steps, the odd steps should be missing \cite{Rokhinson12},
but the experiment actually observes only one missing odd step.
This simple picture of missing odd Shapiro steps, although physically plausible, may not be complete,
and a complete theory for Shapiro steps in the presence of MZMs has not yet been formulated
(see, however, Ref. \onlinecite{Badiane13}).

\section{`Synthetic' Realization of Topological Superconductors}
\label{sec:synthetic}

Before further discussing experimental probes of Ising anyons, we pause to discuss `synthetic'
realizations of topological superconductors because it will be useful to have concrete device structures in mind when we describe procedures for braiding non-Abelian anyons. `Synthetic' systems are important because
there is no known `natural' system that spontaneously enters a topological superconducting phase.
The A-phase of superfluid He-3 \cite{Salomaa85} and superconducting Sr$_2$RuO$_4$ \cite{Mackenzie03} are
hypothesized to possess some topological properties, but it is not known precisely how to bring these systems into topological superconducting phases that
support MZMs, nor is it known precisely how to detect and manipulate Majorana zero modes in these systems \cite{DasSarma06a}.
There are also specific proposals for converting ultracold superfluid atomic fermionic gases into topological superfluids \cite{Tewari07a},
but experimental progress has been slow in the atomic systems because of inherent heating problems.
However, topological superconductivity can occur in`synthetic' systems \cite{Fu08,Sau10a,Lutchyn10,Oreg10,Alicea10,Potter10,Potter11}
that combine ordinary non-topological superconductors with other materials, thereby facilitating interplay between superconductivity
and other (explicitly, rather than spontaneously) broken symmetries.

The following single-particle Hamiltonian is a simple toy model for a topological superconducting wire \cite{Kitaev01}
which illustrates how MZMs can arise at the ends of a 1D wire:
\begin{multline}
H = \sum_{i} \Bigl(-t [c^\dagger_{i+1} c^{}_{i} + c^\dagger_{i} c^{}_{i+1}] 
- \mu c^\dagger_i c^{}_{i}  \\
+ \Delta c_i c_{i+1} + \Delta^* c^\dagger_{i+1} c^\dagger_{i} \Bigr)
\end{multline}
Here, the electrons are treated as spinless fermions that hop along a wire composed of
a chain of lattice sites labelled by $i=1, 2, \ldots, N$. It is assumed that a fixed pair field $\Delta = |\Delta| e^{i\theta}$
is induced in the wire by contact with a 3D superconductor through the proximity effect. To analyze this Hamiltonian, it is useful to
absorb the phase of the superconducting pair field into the
operators $c_j$ and then to express them in terms of their real and imaginary parts:
$e^{i\frac{\theta}{2}} c_j = a_{1,j} + i a_{2,j}$,
$e^{-i\frac{\theta}{2}} c^\dagger_j = a_{1,j} - i a_{2,j}$. The operators $a_{1,j}$, $a_{2,j}$ are
self-adjoint fermionic operators -- $a^\dagger_{1,j}= a_{1,j}$, $a^\dagger_{2,j}= a_{2,j}$ -- i.e. they
are Majorana fermion operators. They are (generically) not zero modes since they do not commute
with the Hamiltonian but they enable us to elucidate the physics of this Hamiltonian since it can be written as:
 \begin{multline}
H = \frac{i}{2}\sum_{j} \bigl[ -\mu a_{1,j} a_{2,j} + (t+|\Delta|) a_{2,j} a_{1,j+1}\\
+ (-t+|\Delta|) a_{1,j} a_{2,j+1}\bigr]
\end{multline}
Now, it is clear that there is a trivial gapped phase (an atomic insulator) centered about the point $|\Delta|=t=0$, $\mu<0$.
The Hamiltonian is a sum of on-site terms $i |\mu| a_{1,j} a_{2,j}/2$, each of which has eigenvalue $-|\mu|/2$ in the
ground state, with minimum excitation energy $|\mu|$. However, there is another gapped phase that includes the points
$t=\pm |\Delta|$, $\mu=0$. At these points, the Hamiltonian is a sum of commuting terms, but they are not on-site.
Consider, for the sake of concreteness, the point $t=|\Delta|$, $\mu=0$. Then the Hamiltonian couples
each site to its neighbors by coupling $a_{2,j}$ to $a_{1,j+1}$. As a result, we can form a set of independent
two-level systems on the links of the chain. Each link is in its ground state $i a_{2,j} a_{1,j+1} = -1$. However,
there are "dangling" Majorana fermion operators at the ends of the chain because $a_{1,1}$ and $a_{2,N}$
do not appear in the Hamiltonian. They are Majorana zero mode operators:
\begin{equation}
\{a_{1,1}, a_{2,N}\} = [H,a_{1,1}] = [H,a_{2,N}] = 0
\end{equation}
If we move away from the point $t=|\Delta|$, $\mu=0$, $a_{1,1}$ and $a_{2,N}$ will appear in the Hamiltonian
and, as a result, they will no longer commute with the Hamitonian. However, there will be a more complicated
pair of operators that are exponentially-localized at the ends of the chain and satisfy Eq. (\ref{eqn:MZM-real-def}).
Thus, the 1D toy model describes a system with localized zero-energy Majorana excitations at the wire ends,
which serve as the defects.

Very similar ideas hold in 2D \cite{Volovik99,Read00}, where an $hc/2e$ vortex in a fully spin-polarized
$p+ip$ superconductor supports a MZM. The 1D edge of such a 2D superconductor supports a chiral Majorana fermion:
\begin{equation}
S = \int dx \, dt\, \chi (i\partial_{t} + v\partial_{x}) \chi
\end{equation}
where $\chi(x,t)=\chi^\dagger(x,t)$ and $\{\chi(x,t), \chi(x',t)\} = 2\delta(x-x')$. When an odd number of vortices
penetate the bulk of the superconductor, the field $\chi$ has periodic boundary conditions, $\chi(x,t)=\chi(x+L,t)$,
where $L$ is the length of the boundary. Then, the allowed momenta are $k = {2\pi n}/{L}$ with $n=0,1,2,\ldots$
and the corresponding energies are $E_n = v k$. The $k=0$ mode is a MZM. If an even number of vortices
penetrate the bulk of the superconductor, $\chi$ has anti-periodic boundary conditions, $\chi(x,t)=-\chi(x+L,t)$
and there is no zero mode because the allowed momenta are $k = {(2n+1)\pi }/{L}$. A vortex may be viewed
as a very short edge in the interior of the superconductor,
so that there is a large energy splitting between the $n=0$ mode and the $n\geq 1$ modes.

Although the toy model described above is not directly experimentally relevant, we can realize either a 1D or a 2D topological superconductor in an experiment,
if we somehow induce spinless $p$-wave superconductivity in a metal in which a single spin-resolved
band crosses the Fermi energy. 
This can be done with a Zeeman splitting that is large enough to fully spin-polarize the system,
but superconductivity has never been observed in such a system;
if induced through the superconducting proximity effect, it is likely to be very weak
since the amplitude of Cooper pair tunneling from the superconductor into the ferromagnet would be very small.
However, the surface state of a 3D topological insulator \cite{Moore07,Fu07,Roy09}
has such a band which can be exploited for these purposes\cite{Fu08}.
Moreover, a doped semiconductor with a combination of spin-orbit
coupling and Zeeman splitting leads, for a certain range of chemical
potentials, to a single low-energy branch of the electron excitation spectrum in
both 2D \cite{Sau10a} and 1D systems \cite{Lutchyn10,Sau10b,Oreg10}. 
In the former case, the Zeeman field must generically be in the direction perpendicular to the 2D system.
In the presence of a superconductor, such a Zeeman splitting must be created by proximity to a
ferromagnetic insulator, rather than with a magnetic field. The exception is a system in which
the Rashba and Dresselhaus spin-orbit couplings balance each other \cite{Alicea10}. In 1D, however,
the Zeeman field can be created with an applied magnetic field, thus making a 1D semiconducting nanowire
with strong spin-orbit coupling and superconducting proximity effect particularly attractive as an experimental platform for investigating Majorana zero-modes.
This idea \cite{Lutchyn10,Sau10b,Oreg10} has been adapted by several experimental groups \cite{Mourik12,Rokhinson12,Deng12,Churchill13,Das12,Finck12}.

\begin{figure}[t]
  \includegraphics[width=2.8cm]{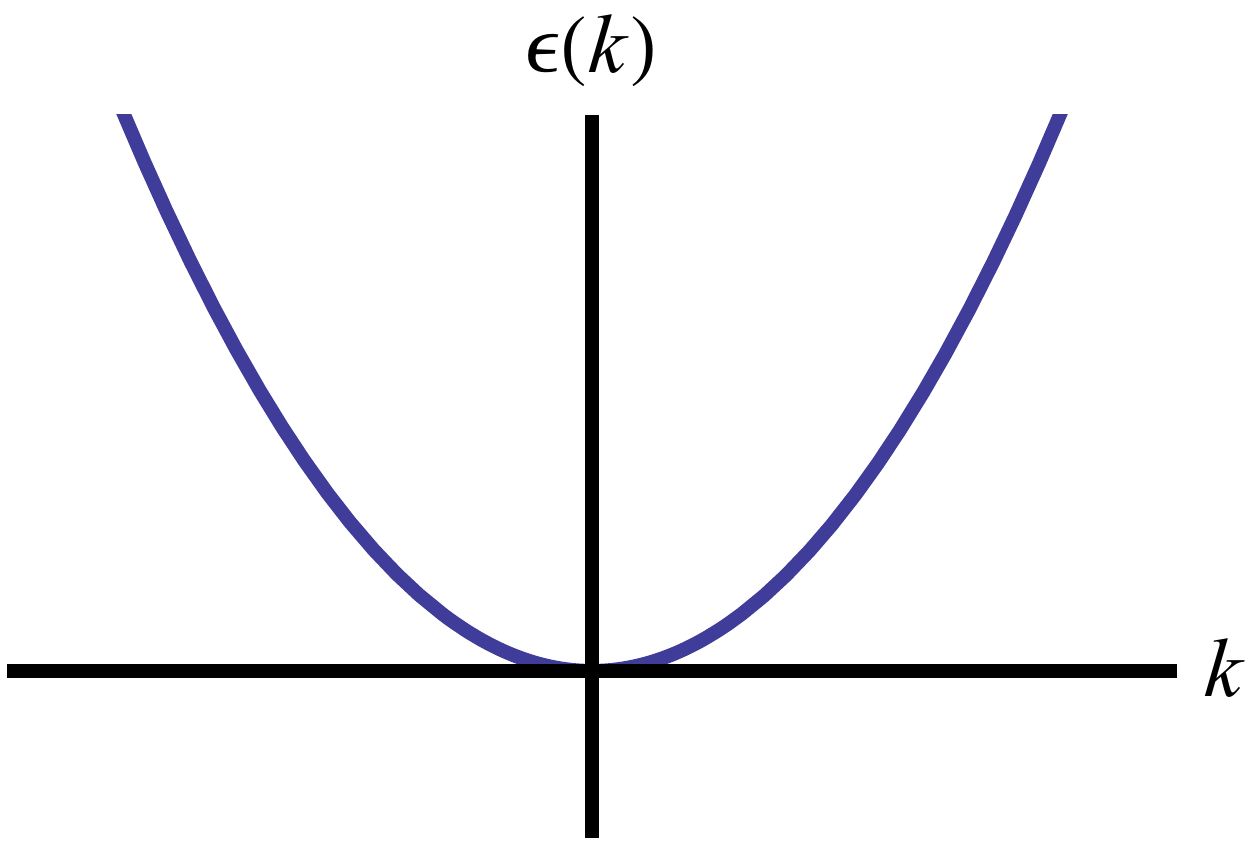}
\includegraphics[width=2.8cm]{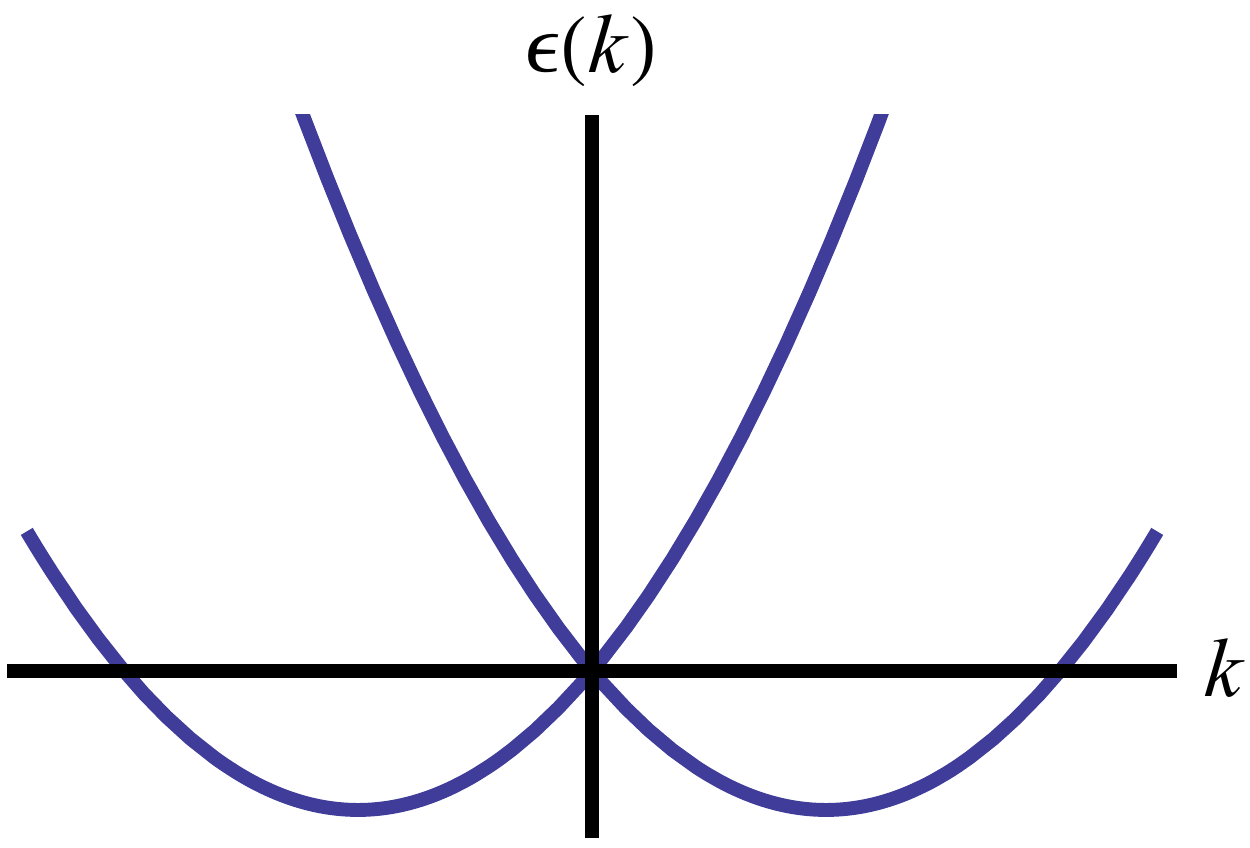}
\includegraphics[width=2.8cm]{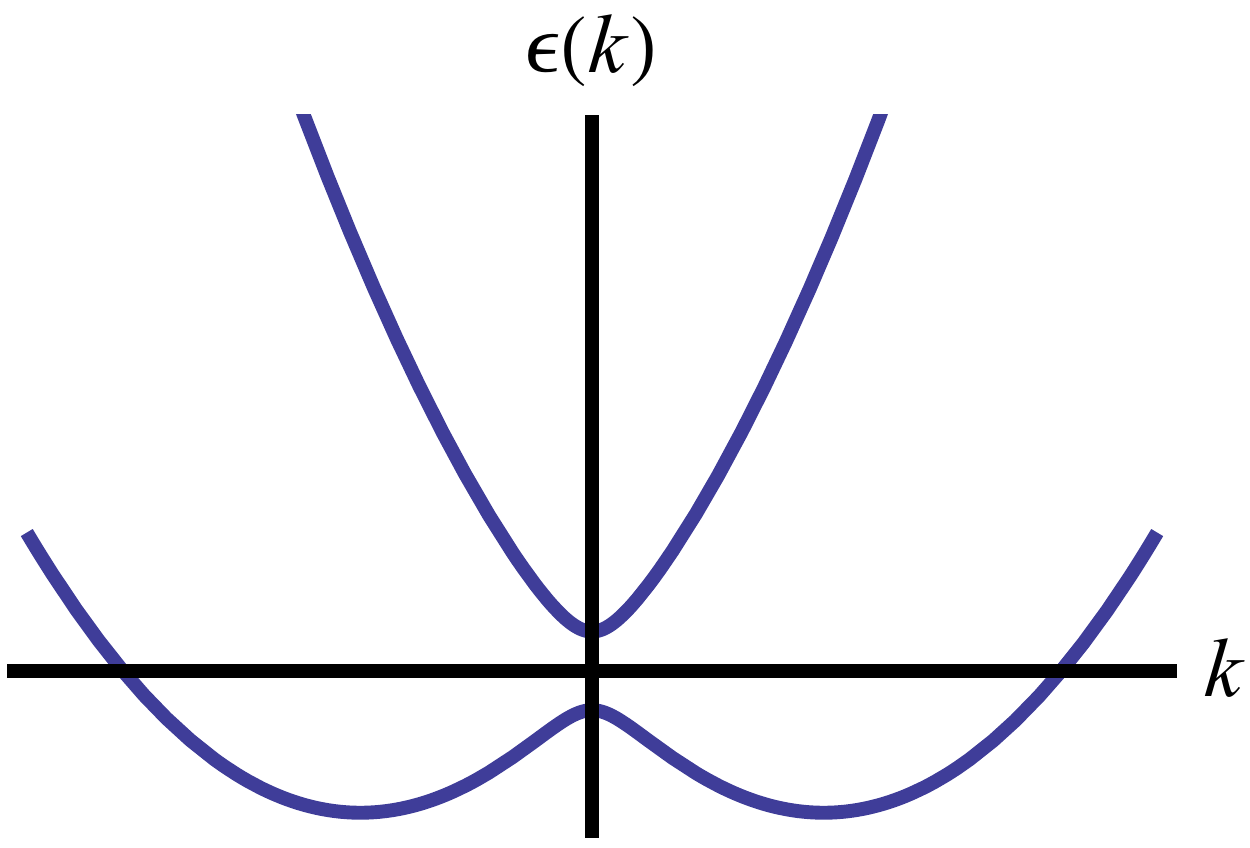}
 \caption{The electron energy $\epsilon(k)$ as a function of momentum $k$ for a 1D wire modeled by the Hamiltonian
in Eq. (\ref{eqn:Lutchyn-model}) for (left panel) vanishing spin-orbit coupling and Zeeman splitting; (center panel) non-zero spin-orbit
splitting but vanishing Zeeman splitting; (right panel) non-zero spin-orbit and Zeeman splitting. In the situation in the right-panel, if
the Fermi energy is close to $\epsilon=0$, then there is effectively a single band of spinless electrons at the Fermi energy.}
  \label{fig:band-picture}
\end{figure}

In all of these cases, the electron's spin is locked to its momentum, rendering it effectively spinless.
Such a situation has the added virtue that an ordinary $s$-wave superconductor can induce
topological superconductivity \cite{Fu08,Zhang08,Sato09b,Sau10a,Sau10b,Lutchyn10,Oreg10}
since the spin-orbit coupling mixes $s$-wave and $p$-wave components.
An effective model for this scenario takes the following form:
\begin{multline}
\label{eqn:Lutchyn-model}
H = \int dx \Bigl[\psi^\dagger \bigl( -\mbox{$\frac{1}{2m}$}\partial_x^2 - \mu
+ i\alpha \sigma_y \partial_x + V_x \sigma_x\bigr) \psi\\
+ \Delta \psi_\uparrow \psi_\downarrow + \text{h.c.}  \Bigr]
\end{multline}
This model is in the topological superconducting phase when the following condition holds \cite{Lutchyn10,Sau10b,Oreg10}:
${V_x}>\sqrt{{|\Delta|^2} + \mu^2}$, i.e. when the Zeeman spin splitting $V_x$
is larger than the induced superconducting gap $\Delta$ and the chemical potential $\mu$ -- a situation
which presumably can be achieved by tuning an external magnetic field $B$
to enhance the Zeeman splitting \cite{Note4}.
(In principle, the system can be tuned by changing the chemical potential as well
using an external gate to control the Fermi level in a semiconductor nanowire, thus adding considerable flexibility to the set up for eventual TQC braiding manipulations of the MZMs.)
When the two sides of this equation are equal, the system
is gapless in the bulk and is at a quantum phase transition between ordinary and topological superconducting phases.
The emergence of an effectively spinless band of electrons in this model is summarized by Fig.
\ref{fig:band-picture}.
Here, for simplicity, we have assumed that there is a single
sub-band, i.e. a single transverse mode, in the wire. If there are more modes, then the requirement is that there must be
an odd number of modes described by Eq. (\ref{eqn:Lutchyn-model}) in the topological
superconducting phase \cite{Lutchyn10,Lutchyn11a,Stanescu11}.
(In addition, there can be any number of modes in the non-topological phase; recall from
Sec. \ref{sec:MZMs-in-SC} that non-topological physics, here in the form of normal bands, may coexist with
the topological bands.)
From the preceding analysis, we see that there is a minimum magnetic field that must be exceeded
in order for the system to be in a topological superconducting phase. In a real system in which
there will be multiple sub-bands, there is a maximum applied magnetic field, too, beyond which the lowest
empty sub-band crosses the Fermi energy.
(Also, at high applied fields, the topological superconducting gap decreases inversely with increasing spin splitting,
thus requiring very low temperatures to study the MZMs \cite{Sau10b}.)
It is important that the magnetic field be perpendicular to the
spin-orbit field. If the latter is in the $y$-direction, as in Eq. (\ref{eqn:Lutchyn-model}), then the applied magnetic
field must be in the $x-z$ plane. In practice, this angular dependence on the magnetic field can be and has been used to study the MZMs in the laboratory \cite{Mourik12}.

\section{Topological Superconductors: Experiments and Interpretation}
\label{sec:experiments}

A number of experimental groups \cite{Mourik12,Rokhinson12,Deng12,Churchill13,Das12,Finck12}
have fabrcated devices consisting of an InSb or InAs semiconductor nanowire
in contact with a superconductor, beginning with the Mourik et al. experiment of Ref \onlinecite{Mourik12}.
Both InSb and InAs have appreciable spin-orbit coupling and large Land\'e $g$-factor so that
a small applied magnetic field can produce large Zeeman splitting. The experiments of Ref. \onlinecite{Mourik12,Churchill13}
used the superconductor NbTiN, which has very high critical field, while the experiments of Refs.
\onlinecite{Deng12,Das12,Finck12} used Al. All of these experiments observed a zero-bias peak (ZBP), consistent with the MZM expectation.
Meanwhile, the experiment of Ref. \onlinecite{Rokhinson12} observed
Shapiro steps in the AC Josephson effect in an InSb nanowire in contact with Nb. 

According to the considerations of the previous two sections, once the magnetic field
is sufficently large that $V_x  >\sqrt{{|\Delta|^2} + \mu^2}$, where $V_x = g \mu_B B$,
the conductance through the wire between a normal lead and a
superconducting one will be $2{e^2}/h$ at vanishing bias voltage and temperature
\cite{Sengupta01,Law09,Fidkowski12,Lutchyn13}, provided that the wire is much longer than
the induced coherence length in the wire (i.e. the typical size
of the localized MZMs). The five experiments of Refs. \onlinecite{Mourik12,Deng12,Churchill13,Das12,Finck12}
observe a zero-bias peak at magnetic fields $B \gtrsim 0.1$ T, provided that the field is perpendicular to
the putative direction of the spin-orbit field. The peak conductance is, however, significantly smaller than $2{e^2}/h$
in all of these experiments. Moreover, the wires
appear to be short, as compared to the inferred coherence length in the wires, raising the question of
why the MZM peak is not split into two peaks away from zero bias voltage
due to the hybridization of the two end MZMs overlapping with each other (although some signatures of ZBP splitting are indeed observed in some of the data
\cite{Mourik12,Deng12,Churchill13,Das12,Finck12}).
In addition, the subgap background conductance is not very strongly suppressed at low non-zero voltages,
i.e. the gap appears to be `soft'. Finally, the appearance of the peak at
$B\sim 0.1$ T does not appear to be accompanied by a closing of the gap, as expected at a quantum phase transtion.

\begin{figure}[t]
  \includegraphics[width=8cm]{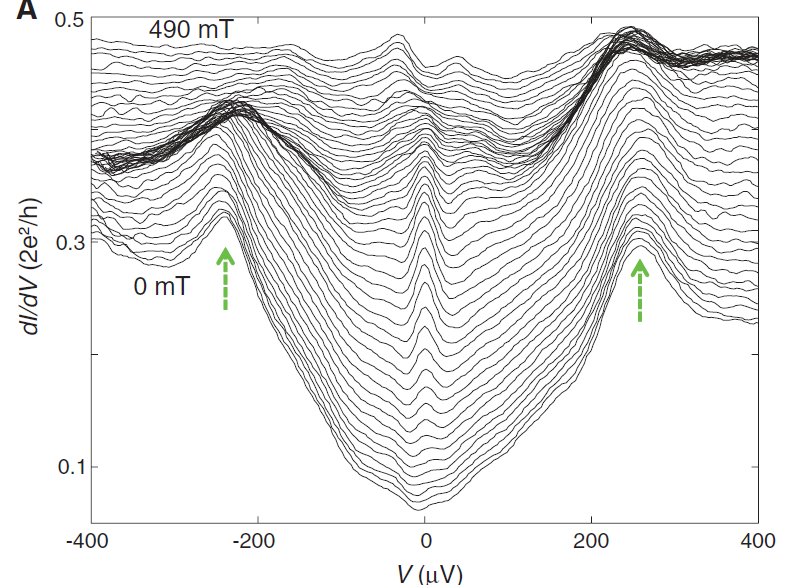}
 \caption{The experimental differential conductance spectrum in an InSb nanowire in the presence of a variable magnetic field showing the theoretically predicted Majorana zero bias peak at finite magnetic field (taken from Ref. \onlinecite{Mourik12}). See the text for a more detailed discussion of the experiment.}
\label{fig:ZBP}
\end{figure}

However, the peak conductance is expected to be suppressed by
non-zero temperature in conjunction with finite tunnel barrier,
and in short wires (see, e.g. Refs. \onlinecite{Stanescu12,LinC12}).
Some of the experiments do appear to find that the zero-bias peak sometimes splits \cite{Das12,Finck12,Churchill13}
and that this splitting oscillates with magnetic field, as predicted \cite{DasSarma12}, although a detailed quantitative comparison between experimental and theoretical zero bias peak splittings has not yet been carried out in depth, and such a comparison necessitates detailed knowledge about the experimental set ups (e.g. whether the system is at constant density or constant chemical potential \cite{DasSarma12}) unavailable at the current time.
The softness of the gap may be due
to disorder, especially inhomogeneity in the strength of the superconducting proximity effect \cite{Takei13b}
or perhaps an inverse proximity effect at the tunnel barriers where normal electrons could tunnel in from the metallic leads into the superconducting wire, leading to subgap states \cite{Stanescu14}.
The softness of the gap may also help explain why the zero-bias conductance is suppressed from its expected
quantized peak value,
although other factors (e.g. finite wire length, finite temperature, finite tunnel barrier, etc.) are likely to be playing a role too.
Very recent experimental efforts \cite{Krogstrup15,Chang15} using epitaxial superconductor (Al)-semiconductor (InAs) interfaces have led
to hard proximity gaps.
The absence of a visible gap closing at the putative quantum phase transition may be due to the vanishing amplitude
of bulk states near the ends of the wire \cite{Stanescu12}; a tunneling probe into the middle
of the wire would then observe a gap closing (but presumably no MZM peaks which should decay
exponentially with distance from the ends of the wires). Such a gap closing has been tentatively identified in the experiments on InAs nanowires in Ref. \onlinecite{Das12}.

In the experiment of Ref. \onlinecite{Rokhinson12}, it was observed that the $n=1$ Shapiro step was suppressed
for magnetic fields larger than $B=2$T. If this is the critical field beyond which $g \mu_B {B_x} = V_x >\sqrt{{|\Delta|^2} + \mu^2}$
in this device, then all of the odd Shapiro steps should be suppressed. However, one could argue that the fermion
parity of the MZMs fluctuates more rapidly at higher voltages so that only the $n=1$ step is suppressed.
More theoretical work is necessary to understand Shapiro step behavior in the presence of MZMs (see, however,
Ref. \onlinecite{Badiane13}).

ZBPs can occur for other reasons, which must be ruled out before one can conclude that the experiments of
Refs. \onlinecite{Mourik12,Deng12,Churchill13,Das12,Finck12} have observed a MZM, particularly since the expected
conductance quantization associated with the perfect Andreev reflection has not been seen. The Kondo effect leads to
a ZBP \cite{LeeEJH12}. In the presence of spin-orbit coupling and a magnetic field, the two-level system may not be the two states
of a spin-1/2, but may be a singlet state and the lowest state of a triplet, which become degenerate at some non-zero
magnetic field \cite{LeeEJH12}. Alternatively, the ZBP may be due to `resonant Andreev scattering'. Of course, a MZM is a type
of resonant Andreev bound state so this alternative really means that there may be an Andreev bound state at the end of the
wire that is not due to topological superconductivity but is `accidentally' (i.e. at one point in parameter space,
rather than across an entire phase) at zero energy.
ZBPs could also arise simply due to strong disorder due to antilocalization at zero energy in 1D systems without time-reversal,
charge conservation, or spin-rotational symmetry, usually called class D superconductors \cite{Bagrets12}.

The multiple observations of a zero-bias peak in different laboratories, occuring only in parameter regimes consistent with
theory \cite{Brouwer11a,Brouwer11b,Lobos12,Crepin14} substantiate these interesting observations in semiconductor nanowires
and show that they are, indeed, real effects and not experimental artifacts.
Although these experiments are broadly consistent with the presence of Majorana zero modes at the ends
of these wires, there is still room for skepticism, which can be answered by showing that the ZBPs evolve as
expected when the wires are made longer, the soft gap is hardened (which has happened recently \cite{Krogstrup15,Chang15}),
and the expected gap closing observed at the quantum phase transition.
Finally, experiments that demonstrate the fractional AC Josephson effect and the expected non-Abelian
braiding properties of MZMs would settle the matter.

Very recently, there has been an interesting new development: the claim of an observation of MZMs in metallic ferromagnetic
(specifically, Fe) nanowires on superconducting (specifically, Pb) substrates where ZBPs appear at the wire ends without the application of any external magnetic field,
presumably because of the large exchange spin splitting already present in the Fe wire \cite{Nadj-Perge14}.  There have been several theoretical analyses
of this ferromagnetic nanowire Majorana platform \cite{Brydon14,Hui14,Li14,Dumitrescu15,Peng15} showing that such a system is indeed generically capable of supporting MZMs without any need for fine-tuning of the chemical potential, i.e. the system is always in the topological phase since the spin splitting $V_x$ is always much larger
than $\Delta$ and $\mu$.  Although potentially an important development, more data (particularly, at lower temperatures, higher induced superconducting gap values, and longer
wires) would be necessary before any firm conclusion can be drawn about the experiment of Ref. \onlinecite{Nadj-Perge14} since the current experiments, which are carried out at temperatures comparable to the induced topological superconducting energy gap in wires much shorter than the Majorana coherence length, only manifest very weak ($3-4$
orders of magnitude weaker than $2{e^2}/h$) and very broad (broader than the energy gap) ZBPs.
If validated as MZMs, this new metallic platform gives a boost to the study of non-Abelian anyons in solid state systems.

\section{Non-Abelian Braiding}

As noted in the introduction, the primary significance of Majorana zero modes is that they are a mechanism for
non-Abelian braiding statistics, arising from their ground state topological quantum degeneracy.
The braiding of non-Abelian anyons provides a set of robust quantum gates with topological protection
(although, of course, this only applies if the temperature is much lower than the energy gap and
all anyons are kept much further apart than the correlation length, so that the system is in the
exponentially-small Majorana energy splitting regime). These braiding properties are also the most direct and
unequivocal way to detect non-Abelian anyons -- including, as a special case, those supporting Majorana zero modes.

It is useful, at this point, to make a distinction between the two computational uses of braiding, for unitary gates and for
projective measurement. Braiding-based gates can operate in essentially the same way for quasiparticles in
a topological phase and for defects in an ordered (quasi-topological) state. However, braiding-based measurement procedures
rely on interferometry, which is only possible if the motional degrees of freedom of the objects being braided are
sufficiently quantum-mechanical. This will be satisfied by quasiparticles at sufficiently low temperatures, but
the motion of defects is classical at any relevant temperature except, possibly, in some special circumstances.

\begin{figure}[t]
  \includegraphics[width=8cm]{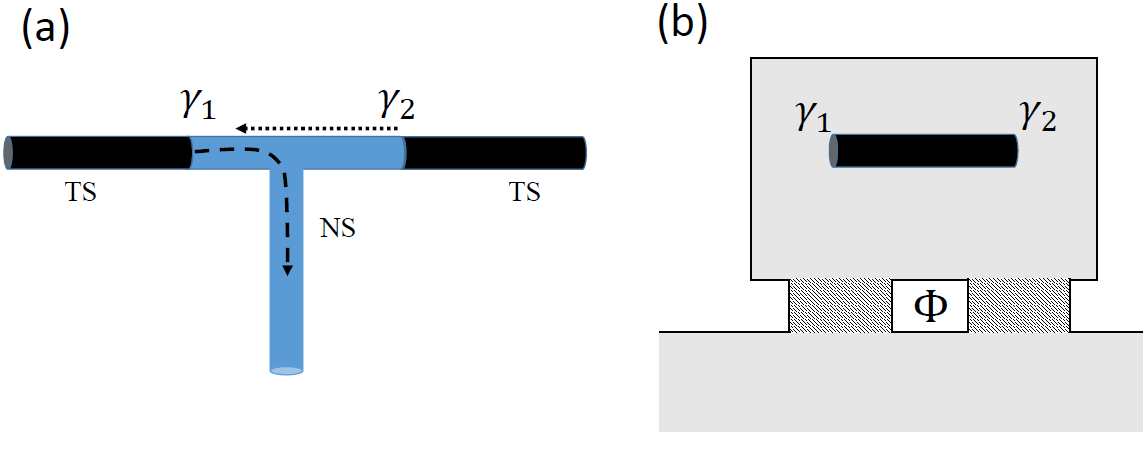}
 \caption{(a) MZMs localized at domain walls between topological superconducting (TS) and normal superconducting (NS) phases
can be moved by tuning regions between these phases to move the domain walls \cite{Alicea11}.
(b) As explained in the text, a measurement-only
scheme can replace actual movement of MZMs. A pair of MZMs can be measured by tuning the flux $\Phi$ through
a SQUID loop to decouple the superconducting island on which the pair resides. This causes the
island and nanowire to be in a superselection sector of fixed electrical charge \cite{Hyart13}.}
\label{fig:moving-MZMs}
\end{figure}

Consider, first, braiding-based gates. As noted above, braiding two anyons that support MZMs (either quasiparticles or defects)
causes the unitary transformation in Eq. (\ref{eqn:braiding-unitary}). But how are we actually supposed to perform the braid?
Here, quasi-topological phases have an advantage over topological phases (which no one has presently proposed to build).
In a true topological phase, it may be very difficult to
manipulate a quasiparticle because it need not carry any global quantum numbers. However, in an Ising-type quantum Hall state,
the non-Abelian anyons carry electrical charge, and one can imagine moving them by tuning electrical gates
\cite{DasSarma05}. In the case of a
2D topological superconductor, MZMs are localized at vortices, and one can move vortices quantum mechanically
through an array of Josephson junctions by tuning
fluxes. In a 1D topological superconducting wire MZMs are localized at domain walls between the topological superconductor and a non-topological
superconductor or an insulator (e.g. at the wire ends). These domain walls can be moved by tuning the local chemical potential or magnetic field.
In short, it is easier to `grab' quasiparticles
when they are electrically-charged and, potentially, easier still to grab a defect when it occurs at a boundary between two phases between
which the system can be driven by varying the electric or magnetic field \cite{Alicea11}.
The latter scenario is exemplified in Fig. \ref{fig:moving-MZMs}a. There are in fact many theoretical proposals on how to braid
the end-localized MZMs using electrical gates in various $T$ junctions made of nanowires,
all of which depend on the ability of external gates in controlling semiconductor carriers.
The potential to manipulate MZMs through external electrical gating is, in fact, one great advantage of semiconductor-based Majorana platforms.

In both cases, quasiparticles and defects, it turns out {\it not} to be necessary to move quasiparticles to braid them. Instead, 
one can effectively move non-Abelian anyons via a ``measurement-only'' scheme \cite{Bonderson08b,Bonderson08c}.
Through the use of ancillary EPR pairs and a sequence of measurements, quantum states can be teleported from one qubit to another.
Similarly, a measurement involving an ancillary quasiparticle-quasihole or defect-anti-defect pair can be used to teleport
a non-Abelian anyon. A sequence of such teleportations can be used to braid quasiparticles. The required sequence of
measurements can be performed without moving the anyons at all, as illustrated by the flux-based scheme of
Refs. \cite{Akhmerov09,Hassler10,Hyart13}. By tuning Josephson couplings (which can be done by varying the flux through SQUID loops),
pairs of MZMs can be measured electrostatically, as depicted in Fig. \ref{fig:moving-MZMs}b.
The fermion parity of a pair of MZMs is measured by isolating that pair on a small superconducting island so that
the two parity states differ by an electrostatic charging energy. When the Josephson coupling between the island a large superconductor is non-zero,
that pair of MZMs is not measured, and a different pair (possibly involving one member of the first pair of MZMs) can be measured.
Thereby, a measurement-only braiding scheme can be implemented without moving any defects at all; all that is necessary is to teleport
their quantum information.

\begin{figure}[t]
  \includegraphics[width=4cm]{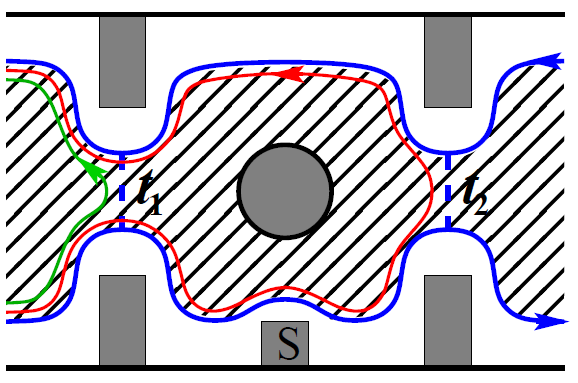}
\hskip 0.2 cm
\includegraphics[width=4cm]{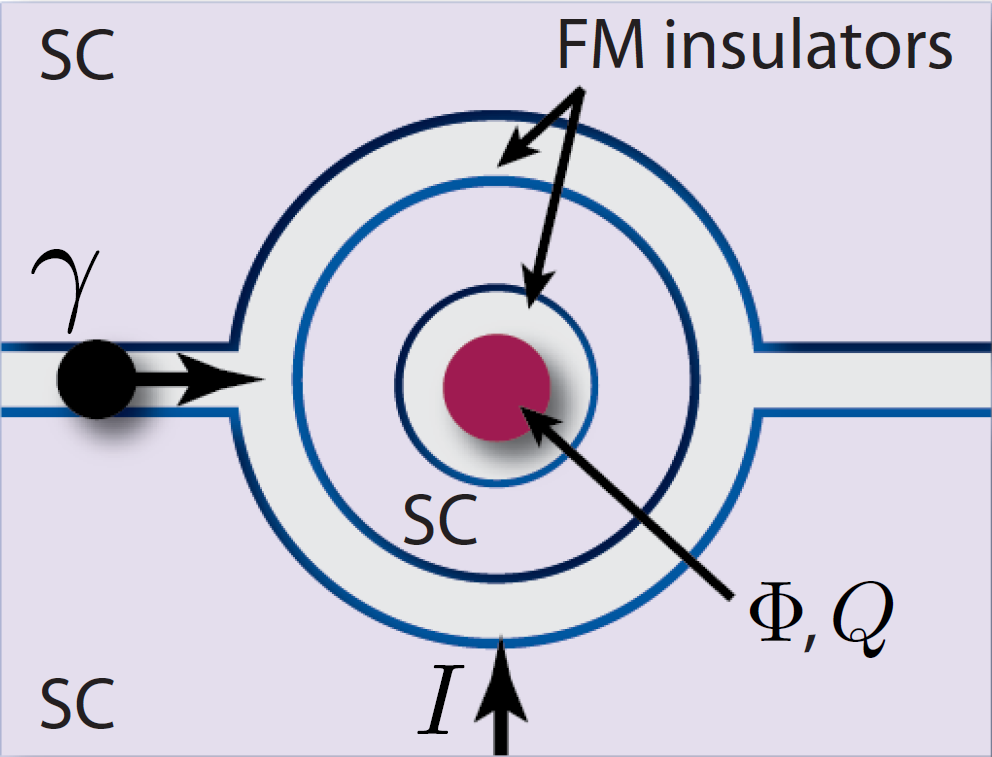}
 \caption{(Left panel) With a two-point contact interferometer in a quantum Hall state, it is possible to detect topological charge
and, thereby, read-out a qubit by measuring electrical conductance (taken from Ref. \onlinecite{Nayak08}.
(Right panel) In a long Josephson junction with two arms, different paths for Josephson vortices can interfere,
thereby enabling the detection of topological charge through electrical measurement (taken from Ref. \onlinecite{Alicea12a}).}
\label{fig:interferometers}
\end{figure}

The second use of braiding is for interferometry-based measurement. This can only be done when the non-Abelian anyons are ``light" so that two different braiding paths can be interfered. This can be done with charge $e/4$ quasiparticles in Ising-type
$\nu=5/2$ fractional quantum Hall states. The two point contact interferometer depicted in Fig. \ref{fig:interferometers}a measures the ratio
between the unitary transformations associated with the two paths. In the case of non-Abelian anyons, this is not merely a phase.
For Ising anyons, there is no interference at all when an odd number of MZMs is in the interference loop. When an even number is
in the interference loop, the interference pattern is offset by a phase of $0$ or $\pi$, depending on the fermion parity of the
MZMs in the loop. The experiments of Refs. \onlinecite{Willett09,Willett10,Willett13a,Willett13b} are consisent with these
predictions, but their interpretation has been questioned \cite{Rosenow09}.

Domain walls in nanowires are always classical objects whose position is determined by gate voltages. Abrikosov
vortices in 2D topological superconductors are similary classical in their motion. However, Josephson vortices, whose cores lie in the insulating
barriers between superconducting regions, may move quantum mechanically, thereby making possible an interferometer such as that
depicted in Fig. \ref{fig:interferometers}. Moreover, the fermionic excitations at the edge of
a superconductor are light and can be used to detect the
presence or absence of a MZM (but not to detect the quantum information encoded in a collection of MZMs).

\section{Quantum Information Processing with Majorana Zero Modes}
\label{sec:quantum-inf}

There are two primary approaches to storing quantum information in MZMs: ``dense'' and ``sparse'' encodings.
In the dense encoding, $n$ qubits are stored in $2n+2$ MZMs $\gamma_{1}, \gamma_{2}, \ldots, \gamma_{2n+2}$.
The two basis states of the $k^{\rm th}$ qubit correspond to the eigenvalues $i \gamma_{2k-1} \gamma_{2k} = \pm 1$.
The last pair, $\gamma_{2n+1}, \gamma_{2n+2}$ is entangled with the total fermion parity of the $n$ qubits so that the
state of the system is always an eigenstate of the total fermion parity of all $2n+2$ MZMs. The advantage of this encoding is
that it is easy to construct gates that entangle qubits. The disadvantage is that the last pair of MZMs is always highly entangled with
the rest of the system, so errors in that pair (even if rare) can infect all of the qubits. In the sparse encoding,
$n$ qubits are stored in $4n$ MZMs $\gamma_{1}, \gamma_{2}, \ldots, \gamma_{4n}$. For all $k$,
we enforce the condition $\gamma_{4k-3} \gamma_{4k-2} \gamma_{4k-1} \gamma_{4k}=-1$, i.e. the total fermion
parity of the set of four MZMs is even in the computational subspace. The two basis states of the $k^{\rm th}$ qubit correspond
to the two eigenvalues $i \gamma_{4k-3} \gamma_{4k-2} = \pm 1$. (Note that, in the computational subspace,
$i \gamma_{4k-3} \gamma_{4k-2} = i \gamma_{4k-1} \gamma_{4k}$.) Since each quartet of MZMs has fixed fermion
parity, it is easier to keep errors isolated. However, there are no entangling gates resulting from braiding alone. In order
to entangle qubits, we need to perform measurements in order to pass from one encoding to the other.

The gates $H, T, \Lambda(\sigma_{z})$ form a universal gate set, where $H$ is the Hadamard gate, $T$ is the $\pi/8$-phase
gate, and $ \Lambda(\sigma_{z})$ is the controlled-Z gate:
\begin{equation}
H = \frac{1}{\sqrt{2}}\begin{pmatrix}
1 & 1 \\1 & -1
\end{pmatrix},\,
T = \begin{pmatrix}
1 & 0 \\ 0 & e^{i \pi/4}
\end{pmatrix},\,
Z = \begin{pmatrix}
1 & 0 \\ 0 & -1
\end{pmatrix}.\nonumber
\end{equation}
In order to apply the Hadamard gate to
the $k^{\rm th}$ qubit, we perform a counter-clockwise exchange of the MZMs $\gamma_{4k-2}$
and $\gamma_{4k-1}$. In order to apply $\Lambda(\sigma_{z})$ to two qubits encoded in 8 MZMs,
we first change to the dense encoding, in which the two qubits are encoded in $6$ MZMs. This involves a measurement.
In this encoding, a braid implements $\Lambda(\sigma_{z})$. Finally, we introduce an ancillary pair of MZMs
and perform a measurement in order to return to the sparse encoding. To be more precise, suppose that
our two qubits are associated with MZMs $\gamma_{1}, \ldots \gamma_{8}$ in the sparse encoding, with the first
four encoding the first qubit and the second four the second qubit. First, we measure $i\gamma_{4}\gamma_{5}$.
If it is equal to $+1$, then the remaining MZMs form a dense encoding of the two qubits. If the measurement returns
$-1$, a straightforward correction will be needed. Then we perform a counter-clockwise
exchange $3$ and $6$ (which are the middle two of the remaining MZMs) followed by clockwise exchanges of $1$ and $2$
and of $7$ and $8$. Finally, we return to the sparse encoding by introducing an ancillary pair of MZMs,
which we will call $\gamma_{4}$ and $\gamma_{5}$, which are in the known state $i\gamma_{4}\gamma_{5}=1$.
Then a measurement of $\gamma_{5}\gamma_{6}\gamma_{7}\gamma_{8}$ returns the system to the sparse encoding.

A single-qubit phase gate can be performed by bringing two MZMs close together for a period of time, $t$,
so that their two states will be split in energy by $\Delta E$, and then pulling them apart again:
\begin{equation}
U = \begin{pmatrix}
1 & 0 \\ 0 & e^{i \Delta E t}
\end{pmatrix}
\end{equation}
This is a completely unprotected operation. Topology does not help us here. If we had perfect control over our system,
then we would be able to control $\Delta E$ and $t$ precisely so that we could set $\Delta E t = \pi/4$ and
obtain a $T$ gate. (Indeed, this is the type of control on which ``conventional'' qubits rely.) However, we do not expect
to have such perfect control, so some error correction will be needed. 
In the case of the $T$ gate, for example, we can use ``magic state distillation" \cite{Bravyi05} to provide a higher fidelity $T$ gate.
Fortunately, the availability of topologically-protected
operations, namely protected Clifford operations, to perform error correction and distillation means fewer physical qubits should be required in the topological case compared to the conventional case.

The basic idea behind distillation is as follows. If we can produce the state
$|a\rangle =|0\rangle + e^{i\pi /4}|1\rangle$ on demand, this is as good as being able to apply the $T$ gate
since we can perform a CNOT gate with $|a\rangle$ as the control qubit and our data qubit as the target. This is
followed by a measurement of the latter and a correction by a Clifford operation if the measurement returns a $+1$.
Therefore, the goal is to produce a high fidelity copy of $|a\rangle$. This can be done in a variety of ways and has become, now, highly optimized
\cite{Paetznick13,Jones13,Eastin12,Bravyi12,Meier12}.

The original distillation protocol \cite{Bravyi05,Bravyi06} proceeds by taking 15 approximate copies of $|a\rangle$:
$|{\tilde a}_{1}\rangle, \ldots, |{\tilde a}_{15}\rangle$, each with fidelity at least $1-\varepsilon$. The tensor product
of these 15 states is projected on the code subspace of the $[[15,1,3]]$ Reed-Muller code. This stabilizer code has the following properties: it
encodes 1 logical qubit in 15 physical qubits; it can detect up to 2 phase ($Z$) errors and up to 6 bit ($X$) errors; and, remarkably, the logical state
$|a\rangle$ is the product of 15 copies of $|a\rangle$. Consequently, given 15 noisy copies of $|a\rangle$,
we can check 14 stabilizers to see if it is consistent with being in the Reed-Muller code subspace. If it is, we can decode the
resulting 15 physical qubits into a logical qubit, which will be a purified version of the state $|a\rangle$,
with fidelity $1-\varepsilon_{out} \approx 1-35\varepsilon^3$, in the limit that $\varepsilon$ is small.
Distillation improves the fidelity so long as the initial fidelity $\varepsilon$ exceeds the threshold found by solving
$\varepsilon_{out}(\varepsilon) = \varepsilon$.  The threshold is roughly $\varepsilon_0\approx 0.141$
\cite{Bravyi06}.
The distillation protocol can be applied recursively to achieve even higher fidelities on the state $|a\rangle$.
Practically, the fidelity of the Clifford operations implementing the stabilizer checks dictates the minimum $\epsilon_{out}$ achievable using the distillation protocol.  For example, to achieve $\epsilon_{out}\approx 10^{-12}$, a reasonable value for quantum algorithms, the Clifford operations must also have fidelity of $10^{-12}$ \cite{Brooks13}.  Conventional qubit systems will require, e.g., the surface code to achieve such fidelities on the Clifford operations, while topological qubit systems may achieve this fidelity naturally.
Thus, a potential advantage of MZM-based TQC would be the need for fewer qubits and fewer gate operations than in conventional quantum computation.

A given quantum algorithm must be decomposed into a circuit consisting of gates drawn from a fault-tolerant universal gate set, such as the set consisting of $H, T, \Lambda(\sigma_{z})$.
Quantum algorithm decomposition methods based on algebraic number theory have recently dramatically reduced the number of $T$ gates required to implement a given quantum algorithm \cite{Selinger12,Kliuchnikov12,Ross14}.
By additionally allowing an ancilla qubit and measurement to be used during decomposition, another constant factor reduction in the number of $T$ gates can be achieved \cite{Paetznick14,Bocharov14a,Bocharov14b}.
The latter technques are referred to as probabilistic ``Repeat-until-Success" (RUS) circuits.
These aforementioned methods, as well as, e.g., techniques to produce Fourier angle states \cite{DuclosCianci14}, may be ultimately hybridized to more efficiently and fault-tolerantly implement a quantum algorithm using Majorana anyons.

Before concluding this section, we briefly mention some of the potential problems in carrying out TQC with the current Majorana nanowire systems.  First, the soft gap problem alluded to above indicates the presence of considerable non-thermal subgap fermionic states which would cause 'quasiparticle poisoning' of the MZM as the Majorana will hybridize with the subgap fermions and decay (and thereby lose its non-Abelian anyonic character).  Thus, poisoning by stray subgap non-thermal quasiparticles puts an absolute upper bound on the effective Majorana coherence time since poisoning will directly destroy the fermion parity at the heart of the proposed non-Abelian TQC.  Recent experimental work has suppressed  quasiparticle poisoning considerably, leading to possible coherence times as long as 1 minute
\cite{vanWoerkom15,Higginbotham15}.  Another issue is that the current experimental topological gap is rather small (a few K) whereas the Majorana splitting due to the overlap of the MZMs from the two ends of the nanowire are likely to be in the range of 100-200 mK (since the current nanowires are rather short).  The lack of a large separation between these two energy scales introduces complications since the TQC braiding operations must be slow ("adiabatic") compared with the topological gap energy and fast (so that one is in the topologically protected regime) compared with the Majorana splitting energy.  Improvement in materials should lead to larger (smaller) gap (splitting), making this issue go away eventually.  Finally, the current ZBPs, even assuming that they are indeed the predicted MZM conductance peaks, are much smaller (by more than an order of magnitude) than the quantized MZM conductance value of $2{e^2}/h$ associated with the Majorana-induced perfect Andreev reflection, perhaps because of finite temperature, short wire length, and finite tunnel barrier at the interfaces.  This could lead to severe visibility problem during Majorana braiding with very weak signal to noise ratio, necessitating considerable measurement averaging.  Only future braiding experiments could actually decisively establish whether the observed ZBPs in the nanowire tunneling measurements are indeed the predicted MZMs or not.

\section{Outlook}

It does not seem fanciful to compare Majorana systems and nonabelian topological quantum systems in
general with the field effect  transistor (FET). Both are sweet theoretical solutions to the problem of
efficient processing of signals and the information they carry. (For FETs, of course, this theoretical solution has turned out,
through Moore's law, to be an astounding practical engineering success as well, leading to the modern IT universe we live in.)
The kinds of information (classical versus quantum) and the energy scales (eV versus meV) are different,
just as the two ideas are temporally separated by more than 50 years, but each proposes a radical solution to an information processing roadblock.
In each case, the roadblock was not absolute but sufficiently daunting to inspire serious and sustained effort.
There were pre-transistor electronic computers, and it may well be possible to build a pre-topological quantum computer
through an extraordinary investment in error correction using ordinary non-topological qubits \cite{Nielsen00}.
As with our current efforts to build Majorana zero mode systems,
the history of the FET was anchored in materials development and required a rethinking of solid state physics
(involving substantial and continuous developments in surface science, semiconductor physics, materials growth, and lithography).
Today, building topological materials will push the frontiers of purity and precision in materials growth
and force us to extend our ability to model exotic bulk materials, interfaces, and, finally, devices.
Since our entire civilization now turns around the transistor, it would be grandiloquent to claim
any untested technology as the new transistor, and we make no such claim. No one can see the future.
However, we have arrived at a gateway where, in the next few years, our ability
to process information may explode disruptively; there is certainly a large heterogeneous international
effort in this direction of building quantum information processing devices and circuits.
In such a world the topological route is the analog of the FET.

Edgar Lilienfeld filed the first FET patent in 1925. It was in an entirely metallic system in which
the required electronic depletion was too difficult to accomplish reliably. It took roughly four decades and the advent of
semiconductor devices to realize the initial FET vision. Where do we stand with Majorana zero mode systems today?
Experimentalists have picked the most promising materials: high Land\'e $g$-factor (to keep the applied $B$-fields moderate), high spin-orbit coupling (to
strongly lock the spin and momentum bands in order to produce a large topological superconducting gap),
low Schottky barriers and good epitaxial contact (to facilitate induced superconductivity), and high mobility (for coherent transport),
among what was known, i.e. “lying around”, and predicted by the theorists.  Incremental improvements in nanowire design, pacification of interfaces,
and transparency to contacting superconductors, may take us into the regime of workable devices - the transistor of
the 1950s.
But one may expect now that the concepts are clear, that systematic study of materials and their growth and
interface properties could easily lead to new choices. A lesson already emerging from experiments in Copenhagen \cite{Krogstrup15}
can radically reduce subgap states. Their data shows a remarkably crisp BCS spectrum in
epitaxially-coated nanowires \cite{Chang15}.
We cannot of course be sure that the appropriate materials for the future TQC devices have already been developed-- after all, the first transistors were made of
germanium although silicon now rules the electronics world-- but there is now a clear path for progress toward the eventual building of TQC using Majorana anyons.

The “materials frontier” discussed above addresses fidelity and lifetimes – the “numerator” of the expression defining computational power.
The denominator is the clock rate. In the case of a Majorana zero mode system, the key time scale is that of measurement. 
As explained in Section \ref{sec:quantum-inf}, measurement of fermion parity is essential to the distillation of magic states,
and is the leading candidate even for braiding operations. To compute well we must be able to measure quickly
and accurately. The two figures of merit are in fact related: if we can make $n$ measurements within the qubit lifetime,
it does us no good if the fidelity is less than $1-1/n$, for with less fidelity the qubit state will be forgotten long before we make the
$n^{\rm th}$ measurement. For computations in parallel (as will be the norm), the demands on fidelity are proportionately greater because the appropriate $n$ is the {\it total} number of measurements during the computation, not the number on any particular qubit. This tells us that there will be a second  “measurement  frontier”  in which accuracy and speed will be the figures of merit. The leading measurement ideas today involve coupling to superconducting qubits living in an optical cavity and using a shift in the resonant frequency of the microwaves to read out fermion
parity \cite{Hassler10,Hyart13}. This is certainly a good starting point, but the typical number are photon frequencies $\sim 6$ GHz and, with beat frequencies recording the energy spitting of tens of MHz, read out would be limited to perhaps a
MHz clock speed. The inherent energy scales of present Majorana zero mode systems are on the order of $1$ K $\approx 20$ GHz so
there is room to do much better. In fact, to combine the two frontiers
one might envision exploiting exotic superconductors with very large ($\sim 100$K) energy gap, pnictides or cuprates \cite{Kim12,Takei13a}, in conjuction with semiconductor wires to increase the gap protecting Majorana systems and clock rates by an order of magnitude.

Lifetimes/clock rate are hardware specs, but equally important is the scaling of the algorithms that we will run.
There have been roughly three epochs: 1) Circa 1982, Feynman \cite{Feynman82} told us that if we could build a quantum computer, its resource requirements would scale in precisely the same way as the quantum mechanical problems,e.g., quantum chemistry problems, we wished to solve  - replacing the exponential scaling of a classical computer (in which memory must double to account for each new spin-$1/2$ degree of freedom).
2) In the 1990s and 2000s, many key quantum algorithms were developed, including Shor's factoring algorithm
\cite{Shor94}, and a detailed analysis of Feynman's idea.
3) Recent papers have focused on realistic regimes for quantum chemistry, rather
than asymptotics. A straightforward estimate for gate counts of quantum chemistry Hamiltonians
found that the number of computational steps for near equilibration to the ground state scaled rather disastrously; polynomially by very high powers $\sim 11$ so that to obtain the energy of FeO$_2$ to a milliHartree with a
GHz clock rate would take the age of the universe \cite{Wecker14}.
However, improved estimates \cite{Poulin14}, combined with some algorithmic
improvement \cite{Hastings14}, has this time down now to a few minutes (with the most recent polynomial scaling
$\sim 5$th power).
This is one example; now that quantum computers appear to be increasingly realistic,
computer scientists and physicists will find efficient quantum algorthims for an array of problems.
Many of these will be physical (e.g., quantum field theory \cite{Jordan12}
and many-body localization are attractive targets \cite{Bauer14}),
but even areas distant from physics are seeing quantum advances.
Deep learning has had a dramatic impact on machine learning in the last few years \cite{Hinton06,Collobert08,Bengio09,LeCunn10}, but there is a computational bottleneck: computation of the true gradient of $L$, where $L$ is the ``log-likelihood function", is classically intractable, leading to classical methods that can efficiently only approximate $\nabla L$. In physical terms, $L$ is an entropy of a transverse field Ising model on a union of complete bipartite graphs. It is now known \cite{Wiebe14} that quantum computers may be used to estimate $\nabla L$ efficiently by emulating the corresponding Ising model, which leads to improved deep learning models using a quantum computer.

But when do we get to the analog of the silicon FET? Presumably we will eventually do better than Majorana zero modes.
Even as we anticipate great breakthroughs in the physics and engineering of Majorana systems,
we can anticipate their eventual eclipse by anyonic systems (e.g. Fibonacci) that have topologically-protected universal quantum operation.
For many years, that phrase primarily meant a dense braid group representation. Majorana zero modes mirror the
topological phase associated with SU(2)$_2$ (see, e.g. Ref. \onlinecite{Nayak08} for an explanation of this notation).
Fibonacci anyons are present in SU(2)$_3$ and
have dense braid group representations. Furthermore,  there is a hint of a potential path toward physical realization
\cite{Mong14a} through a combination of fractional quantum Hall effect (at the $\nu = 2/3$ plateau)
and superconductivity. SU(2) and all levels 5 and higher also have dense braiding but seem physically impractical.
SU(2)$_4$ is an anomaly; it is potentially related to metaplectic anyonic systems \cite{Hastings13} with a proposed realization
\cite{Clarke13a,Lindner12,Barkeshli13a,Barkeshli14}, but braiding alone does not furnish
a dense gate set.
However, recent unpublished work \cite{Levaillant14}
has demonstrated that SU(2)$_4$ becomes universal when braiding is combined with interferometric measurement.

We are poised on the brink of a revolution in our ability to control quantum systems. Topological systems, initially Majorana systems, will play a role. How wide the technological impact will be outside of physics in not foreseeable, but we can say that we are standing at a transition – we are about to learn to process information  - to {\it think}, so to speak, - in the manner that we know the universe operates: quantum mechanically. The first steps in this intellectual journey have been taken with the potential realization of MZMs in the
laboratory \cite{Mourik12,Rokhinson12,Deng12,Churchill13,Das12,Finck12}, but we still have a long way to go.


\end{document}